\documentclass[aps,prd,twocolumn,groupedaddress,nofootinbib]{revtex4}

\usepackage{graphicx}
\usepackage{amssymb}

\begin{document}


\title{The effect of the recoil energy window on the results of direct dark matter experiments}

\author{F. Giuliani}
\email[]{fgiulian@unm.edu}
\affiliation{Department of Physics and Astronomy, University of New Mexico, NM, USA}


\date{\today}

\begin{abstract}
The effect of the chosen analysis energy window on the results of a dark matter experiment is exemplified by the curious intersection of the exclusion plots of the XENON10 and the CDMS experiments. After proving that the narrow energy window XENON10 chose to analyze is indeed the cause of such intersection, a method to determine the high-energy extreme of the recoil energy window an experiment should use is obtained.
\end{abstract}


\maketitle



\begin{figure*}
\begin{center}
\includegraphics[width=.74 \linewidth]{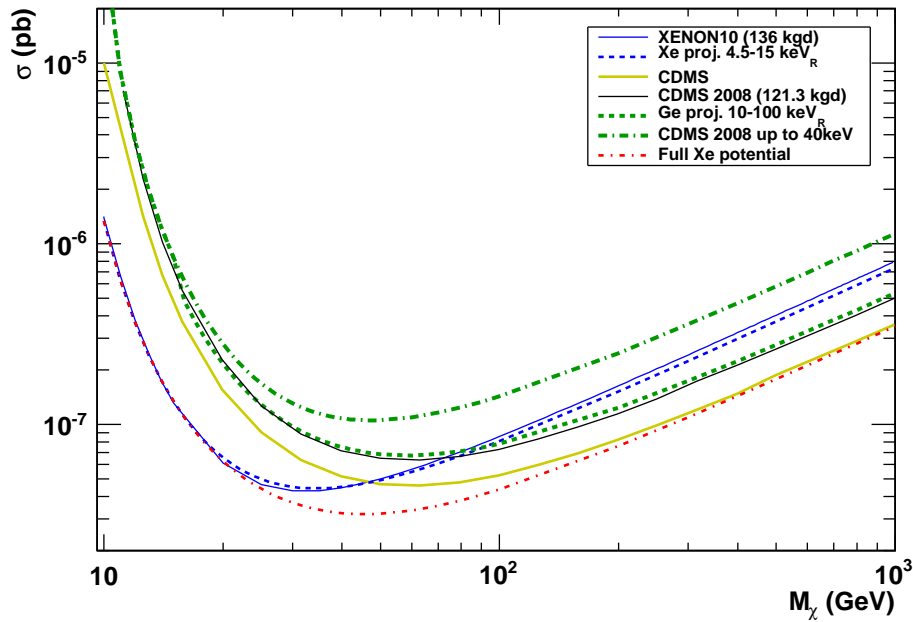}
\caption{Real and calculated exclusion limits of XENON10 and CDMS. The red dash-dotted line is the result XENON10 could have achieved without the observed anomalous events. The green dash-dotted curve shows the effect of a reduced E$_{max}$ on the CDMS 2008 result. The disappearance, in both cases, of the intersection between CDMS 2008 and XENON10 proves that the reduced sensitivity of the latter relative to CDMS at high M$_{\chi}$ is not due to the choice of target (see text).}
\label{excl}
\end{center}
\end{figure*}

As known since decades, the galaxies circular velocities are too high and the gravitational lensing from galaxies and clusters is too strong to be explained by the observed luminous mass. Moreover, the existing cosmological models only explain the observed microwave background anisotropy \cite{wmap} if the dominant component of matter is non-barionic, in concomitance with the prediction by high energy theories beyond the standard model of Weakly Interacting stable and Massive Particles (WIMPs) \cite{mssmPhysRep,kurikamio,scalDM} suitable to constitute such Dark Matter, motivating the many ongoing searches. The current best limits on spin-independent WIMP interactions with nuclei found by the CDMS and XENON10 experiments (solid curves of Fig. \ref{excl}) cross \cite{cdms, weblims} at about 70--80 GeV (2008 CDMS result), intersection that shifts to about 50 GeV when previous CDMS runs are combined \cite{cdms}. Given that the exposures of the CDMS 2008 and XENON10 results are comparable, this seems to imply a lower sensitivity of the XENON10 \cite{xe10} experiment at high WIMP masses, in contradiction with the common knowledge that, on account of the zero-momentum transfer cross section scaling as A$^{2}\mu^{2}$ ($\mu=$ WIMP-nucleus reduced mass), xenon as a target nucleus is more sensitive to spin-independent WIMP scattering than germanium. Although this argument doesn't take into account the faster falloff of xenon's form factor (FF) with respect to germanium, which for sufficiently high recoil energies makes xenon's differential rate fall even below that of Ne \cite{boulhime}, there is another and stronger reason for the above intersection, which becomes apparent when examining the rate equation:

\begin{equation}
\label{rate}
\frac{dR}{dE}=\frac{\rho}{2 M_{\chi}}\frac{\sigma}{\mu^{2}}F(E)\epsilon(E)\int_{v_{min}}^{v_{max}}\frac{f(v)}{v}dv
\end{equation}

\noindent where $\frac{dR}{dE}$ is the rate per unit target mass and recoil energy, $\rho$ is the local halo mass density, $M_{\chi}$ the WIMP mass, $\sigma$ the zero momentum transfer cross section, F(E) the nucleus' FF, $\epsilon(E)$ the detection efficiency, f(v) the halo velocity distribution relative to the detector, v$_{min}$ the minimum velocity an incident WIMP needs to have in order to produce a recoil of energy E and v$_{max}$ the local galactic escape velocity relative to the detector (maximum WIMP velocity in the halo).
While the factor $\frac{\sigma}{\mu^{2}}$ ($\propto A^{2}$) would clearly make xenon about 3 times more sensitive (in terms of $\frac{dR}{dE}$) than germanium, the energy and isotope dependent correction from the FFs (for which both CDMS and XENON10 employed the commonly used Helm FF \cite{helm}) produces, at high momentum transfer, an inversion \cite{fflow}. Based on Eq. (\ref{rate}), the other factors determining $\frac{dR}{dE}$ are the detector independent $\frac{\rho}{M_{\chi}}$, $\epsilon(E)$ and the fraction of the incident WIMP current accessible to the analysis, or velocity acceptance ($\alpha_{v}$), determined by the inverse velocity integral. This integral causes the rate to vanish whenever E is too high, i.e. when $v_{min}(E,A,\mu)\geq v_{max}$.


\begin{figure}
\begin{center}
\includegraphics[width=.9 \linewidth]{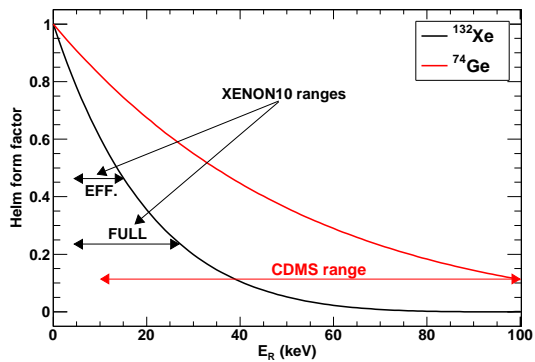}
\caption{Helm FF for $^{132}$Xe and $^{74}$Ge. While germanium's FF is systematically higher than xenon's, in a significant fraction of CDMS's energy range (red arrow) the germanium's FF is actually lower than xenon's FF in the XENON10 energy range (horizontal black arrows). Moreover, the range effectively used by the XENON10 analysis is even narrower than the nominal 4.5--26.9 keV (see text).}
\label{ffs}
\end{center}
\end{figure}


As well known, and reported in  Fig. \ref{ffs} for the most abundant isotopes of natural xenon and germanium \cite{toi}, $^{74}$Ge (35.94\% of natural germanium) and $^{132}$Xe (26.9\% of natural xenon), xenon's FF is generally lower than germanium's. But XENON10 and CDMS use quite different energy ranges: the 4.5 -- 26.9 keV range (black long arrow in Fig. \ref{ffs}) of XENON10 corresponds to a $^{132}$Xe FF range of 0.8--0.24, while CDMS' recoil energy range (red arrow) was 10--100 keV, corresponding, for $^{74}$Ge, to FF values of 0.82--0.11. In particular, above $\sim$70 keV, $^{74}$Ge's FF is lower than $^{132}$Xe's at the right hand extreme of its energy range. Moreover, as will be discussed 
below, the maximum gap method \cite{Yellin} employed by XENON10 in deriving their limits effectively reduces the range of this result to about 4.5--15 keV (short black arrow in Fig. \ref{ffs}), making the FF of 
$^{132}$Xe relevant to the XENON10 result $\gtrsim 0.4$. CDMS's FF is below this value for $E\gtrsim40$ keV, which is more than 50\% of its spectrum. Consequently, the FF difference is significantly less favorable to CDMS, indicating that the reduction in XENON10's sensitivity at high $M_{\chi}$ is primarily due to some other cause.


However, the most convincing argument to discard FF and efficiencies as causes of the reduced XENON10 sensitivity at high M$_{\chi}$ comes from Fig. \ref{excl}, which shows both the CDMS 2008 and the XENON10 results, along with four projections: two reproduce the experimental limits to ensure the correctness of the calculations, while the third displays the limits achievable by a xenon experiment with the same exposure, efficiency and energy threshold of XENON10. The fourth projection shows the limits CDMS 2008 would have with E$_{max}=40$ keV.

In more detail, the green dashed curve is a germanium calculation for the same net exposure and energy interval of CDMS. Based on Ref. \cite{cdms}, CDMS's efficiency was approximated by

\begin{equation}
\epsilon(E)=\left\{\begin{array}{c}24\text{\% if E}< 15 \text{ keV} \\ 
31\text{\% if E} > 15 \text{ keV} \end{array} \right .
\end{equation}

The green dash-dotted calculation differs from the green dashed curve solely by the fact that the recoil energy range is limited to $10\leq E\leq 40$ keV: as in this range germanium's FF is greater than that of xenon in XENON10's range, the disappearance of the intersection between the dash-dotted calculation for CDMS and XENON10 proves the form factors are not responsible for the crossing of the experimental plots.

To reproduce the XENON10 result, it is essential to remember that the maximum gap method \cite{Yellin} used by Ref. \cite{xe10} selects among the intervals (gaps) between any two successive events the one with largest expected number of WIMP events, and extracts the limits from this energy range only. Since the expected rate per unit energy is monotonically decreasing, and the widest gap in Fig. 3 of Ref.  \cite{xe10} is that between the first two events, the energy range actually employed by the XENON10 analysis was about 4.5 -- 15 keV, instead of the full available 4.5 -- 26.9 keV. The dashed blue curve is a calculation for a xenon experiment with the exposure of XENON10 and 4.5 -- 15 keV. The shape of the current XENON10 limits is reproduced sufficiently well by assuming a flat $\epsilon(E)\sim 34.5$\% (which does not include the fiducialization efficiency) and a fiducial raw exposure of 5.4 kg $\times$ 58.6 live days = 316.44 kgd. The net exposure quoted in the legend of Fig. \ref{excl} is taken from Ref. \cite{xe10SD}. The ideal potential of the XENON10 search is instead represented by the red dash-dotted curve, which assumes the same parameters of the blue dashed, except for an energy window of 4.5 keV -- 26.9 GeV, far above the $\sim$2 MeV $^{136}$Xe recoil energy corresponding to a galactic escape velocity (relative to the Milky Way) of 600 km/s and infinite M$_{\chi}$. In particular, since efficiency and FF remained the same, it is clear that the lower than achievable sensitivity of XENON10 at high M$_{\chi}$ is due to the reduced $\alpha_{v}$ determined, via the inverse velocity integral, by the narrow effective energy range \footnote{Note that the CDMS combined exclusions appear to cross the red dashed xenon curve at M$_{\chi}\approx 1$ TeV, due both to the larger exposure of the combination and to the FF effect.}. Since the latter choice is due to the anomalous multiple scattering events reported in Ref. \cite{xe10}, the full xenon potential may be achieved by future larger detectors. In fact, multiple scattering events occasionally deposit part of their energy in the boundary region outside the drift field grid, so that the charge collection is reduced and the secondary scintillation is weaker than it should. Since in larger devices the volume fraction of this boundary region is smaller, future upgrades should detect fewer anomalous events even without possible technical improvements. 


The effect of E$_{max}$ on the analysis' $\alpha_{v}$ can be better understood by considering the kinematic relation yielding the angle $\theta$ between the recoiling nucleus and the incident WIMP direction:

\begin{equation}
\label{vmin}
\cos(\theta)= \frac{q}{2\mu v}=\frac{\sqrt{2EM_{N}}}{2\mu v}
\end{equation}

\noindent where $M_{N}$ is the nucleus' mass.
If the analysis is limited to the recoil energy range [$E_{min},E_{max}$] 
then the visible recoil angles caused by a WIMP of velocity v are constrained by 

\begin{equation}
\label{cosint}
\left\{\begin{array}{l}
\frac{\sqrt{2E_{min}M_{N}}}{2\mu v}\leq\cos(\theta)\leq\frac{\sqrt{2E_{max}M_{N}}}{2\mu v}   \\
0\leq\cos(\theta)\leq1
\end{array}\right .
\end{equation}

Since for nonrelativistic weak scattering the differential cross section is $\propto\cos\theta$ (see, e.g., Ref. \cite{verga}), the $\alpha_{v}$ is given by

\begin{equation}
\label{alfav}
\alpha_{v}=\left\{\begin{array}{l}
1-\frac{\sqrt{2E_{min}M_{N}}}{2\mu v}  \text{      if   } \frac{\sqrt{2E_{max}M_{N}}}{2\mu v}\geq 1\\
\frac{\sqrt{2E_{max}M_{N}}}{2\mu v}-\frac{\sqrt{2E_{min}M_{N}}}{2\mu v} \text{ otherwise}
\end{array}\right .
\end{equation}

\noindent $\alpha_{v}$ is left undefined for $\frac{\sqrt{2E_{min}M_{N}}}{2\mu v}>1$, when the experiment has no sensitivity to WIMPs of velocity v.

\begin{figure}
\begin{center}
\includegraphics[width=.9 \linewidth]{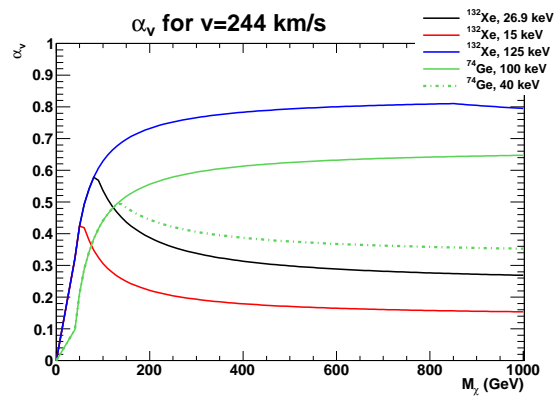}
\caption{$\alpha_{v}$ (Eq. (\ref{alfav})) for WIMPs of velocity 244 km/s incident on $^{132}$Xe and $E_{max}$ of 26.9 keV (black), 15 keV (red) and 125 keV (blue). The curves for $^{74}$Ge and $E_{max}$ of 100 keV (green) and 40 keV (green, dash-dotted) are also shown. The calculations approximated M$_{N}$ with Am$_{p}$, where $m_{p}$ is the proton mass, and employed E$_{min}=4.5$ keV for xenon and 10 keV for germanium.}
\label{maxco}
\end{center}
\end{figure}

More generally, to maximize the acceptance of WIMPs with a given velocity, an experiment needs to have both the lowest possible $E_{min}$ and an $E_{max}$ high enough to ensure that $\frac{\sqrt{2E_{max}M_{N}}}{2\mu v}\geq 1$. Based on these considerations, experiments should try, in analyzing their results, to select $E_{max}$ such that:

\begin{equation}
\label{emax}
E_{max}\geq \frac{2\mu^{2} v^{2}}{M_{N}}=\frac{2M_{\chi}^{2} M_{N}v^{2}}{(M_{\chi}+M_{N})^{2}}
\end{equation}

\noindent for all of their isotopes, the full M$_{\chi}$ range of the limits (or cross sections, for positive signals) to be generated, and velocities up to v$_{max}$.
For illustration purposes, Fig. \ref{maxco} plots $\alpha_{v}$ as a function of M$_{\chi}$ for $^{132}$Xe, and v equal to the macroscopic WIMP wind velocity, i.e., the detector velocity relative to the Milky Way $v_{det}=244$ km/s \cite{lwsm}. While for $E_{max}$ of 26.9 keV $\alpha_{v}$ peaks at M$_{\chi}\approx 85$ GeV, where almost 60\% of such WIMPs are visible to a xenon experiment, for $E_{max}= 15$ keV the peak is reduced to about 42\%, and shifted to an M$_{\chi}$ of 55 GeV. An $E_{max}$ of 125 keV would instead suffice to keep 70--80\% of such WIMPs, in a wide mass range.
The $E_{max}$ of CDMS is instead enough to accept more than 50\% of the WIMPs of mass from 150 GeV up to 1 TeV and velocity of 244 km/s relative to the detector. Had CDMS used E$_{max}=40$ keV, its $\alpha_{v}(244 \text{ km/s})$ would have peaked at M$_{\chi}\approx 150$ GeV to only 50\%. 
Finally, the red (XENON10) and green (CDMS) curves of Fig. \ref{maxco} yield the same WIMP $\alpha_{v}$ at M$_{\chi}\approx 70-80$ GeV, i.e. precisely the region where the black and the blue exclusion plots of Fig. \ref{excl} cross.

But the velocity of the Sun through the Milky Way is well below v$_{max}$, which is determined by the galactic escape velocity relative to the galaxy v$_{esc}$. Estimating the latter velocity is not straightforward, since the work to escape the galaxy depends on the unmeasured size of the Milky Way halo. This size can be guessed from the mass of the halos of galaxies similar to the Milky Way, which can be obtained via gravitational lensing. A frequently employed estimate is v$_{esc}=600$ km/s \cite{lwsm}, corresponding to v$_{max}=844$ km/s.

\begin{figure}
\begin{center}
\includegraphics[width=.9 \linewidth]{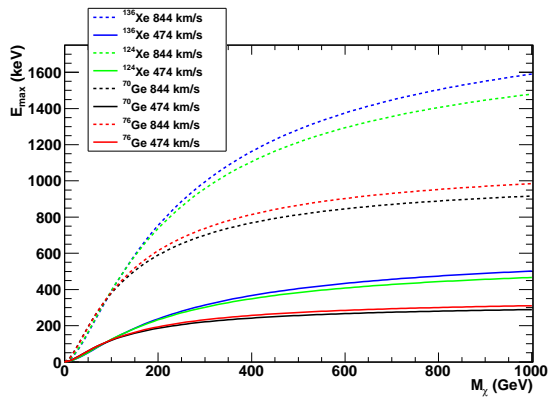}
\caption{E$_{max}$ vs. M$_{\chi}$ for 474 km/s (solid lines) and 844 km/s (dashed lines), and for the lightest and heaviest isotopes occurring in natural germanium and xenon. 474 km/s is the maximum velocity relative to the detector of a WIMP with the most likely velocity relative to the Milky Way.}
\label{emaxfig}
\end{center}
\end{figure}

Fig. \ref{emaxfig} shows the E$_{max}$ values CDMS and XENON should use to retain full WIMP acceptance for M$_{\chi}$ up to 1 TeV. For each experiment, Eq. (\ref{emax}) is plotted for both the lightest and the heaviest of the respective isotopes, and for the velocities of 844 and 474 km/s. For M$_{\chi}=1$ TeV, even CDMS wasn't accepting the whole incident WIMP velocity spectrum. On the other hand, keeping an energy window as wide as suggested by Fig. \ref{emaxfig} for v = 844 km/s (v$_{max}$) and M$_{\chi}$ of 1 TeV would also increase the background acceptance. This circumstance is likely to impose a sacrifice in $\alpha_{v}$ to obtain a significant reduction in background leakage in exchange for a tolerable loss of WIMP acceptance. A velocity region where this can be done is generally near v$_{max}$, i.e., the tail of the WIMP velocity spectrum. Conversely, an experiment should generally try to keep  the velocity region around $\Theta_{0}$, the velocity of an object on a circular orbit of radius equal to the Sun-to-Milky Way center distance R$_{0}$ (for these parameters the values currently recommended by the International Astronomical Union \cite{iau} are 220 km/s and 8.5 kpc, respectively). In fact, only the WIMPs traveling at nearly the Milky Way's rotation velocity dwell for a long time nearby the solar system, the others being on eccentric orbits which will bring them to spend most of the time far from the detector location. This argument generalizes the well known fact that if the velocity distribution with respect to the Milky Way is maxwellian the most likely velocity is exactly the galaxy's circular rotation velocity. Though generally assuming the velocity distribution to be maxwellian, it is common practice to adopt a most likely velocity relative to the galaxy of 230 km/s (making the assumed distribution only quasi-maxwellian), which corresponds to velocities relative to the detector between 0 and 474 km/s\cite{lwsm}. The solid curves of Fig. \ref{emaxfig} for 474 km/s show the E$_{max}$ values needed to retain all of the WIMPs with the most likely velocities: at M$_{\chi}=1$ TeV CDMS needs an E$_{max}$ of about 311 keV, 3 times larger than currently adopted, while XENON requires 0.5 MeV. At M$_{\chi}=0.1$ TeV the required E$_{max}$ values become near, but still somewhat above the 100 keV of CDMS's E$_{max}$.


In synthesis, the above discussion shows, through the comparison of the two currently leading dark matter experiments, the effect of the upper extreme of an experiment's energy window on its final (exclusion or positive) results. For each WIMP velocity and mass, there is a composition dependent E$_{max}$ below which some of the WIMPs above threshold are lost to detection, causing a reduced WIMP acceptance if  E$_{max}$ is chosen too small. In the case of XENON10 this effect is so dramatic to allow CDMS to overcome the larger sensitivity of a heavier target notwithstanding the fact that at the low momentum transfer of the XENON10 result the known \cite{fflow} form factor penalty is eliminated. Although the ideal choice of E$_{max}$ such that all target isotopes are sensitive to all WIMP velocities is both complicated by the uncertainty on the Milky Way's escape velocity and made impractical by the necessity to limit the background acceptance also by reducing the energy range, it should be taken into account when deciding the analysis energy window.

\begin{acknowledgments}
I wish to thank TA Girard for the inspiring discussion which drove my attention to the topic of this paper.\end{acknowledgments}


\end{document}